\documentclass[fleqn,usenatbib,useAMS]{mnras}

\usepackage[dvipsnames]{xcolor}
\usepackage{tikz,hyperref}

\definecolor{lime}{HTML}{A6CE39}
\DeclareRobustCommand{\orcidicon}{
	\begin{tikzpicture}
	\draw[lime, fill=lime] (0,0) 
	circle [radius=0.13] 
	node[white] {{\fontfamily{qag}\selectfont \tiny ID}};
	\draw[white, fill=white] (-0.0625,0.095) 
	circle [radius=0.007];
	\end{tikzpicture}
	\hspace{-2mm}
}

\foreach \x in {A, ..., Z}{\expandafter\xdef\csname orcid\x\endcsname{\noexpand\href{https://orcid.org/\csname orcidauthor\x\endcsname}
			{\noexpand\orcidicon}}
}

\usepackage{newtxtext,newtxmath}
\usepackage{CJK}

\usepackage{booktabs}

\usepackage[T1]{fontenc}
\usepackage{ae,aecompl}

\usepackage{graphicx}	
\usepackage{amsmath}	
\usepackage{booktabs}
\usepackage{times}
\input{epsf}

\title[A hidden WD in $\zeta$ Tau]{Ti\={a}ngu\={a}n ($\zeta$ Tau) as a binary system consisting of a Be-star and an accreting White Dwarf: opening a gate to understanding enigmatic $\gamma$ Cas analogues}

\author[J.~A.~Toal\'{a}, L.~M.~Oskinova \& D.~A.~Vasquez-Torres]{Jes\'{u}s~A.~Toal\'{a}\thanks{E-mail:\,j.toala@irya.unam.mx}$^{1}\orcidA$, Lidia~M.~Oskinova$^{2}\orcidC$ and Diego~A.~Vasquez-Torres$^{1}\orcidB$\\
$^{1}$Instituto de Radioastronom\'{i}a y Astrof\'{i}sica, Universidad Nacional Aut\'{o}noma de M\'{e}xico, 58089 Morelia, Michoac\'{a}n, Mexico\\
$^{2}$Institut f\"{u}r Physik und Astronomie, Universit\"{a}t Potsdam, Karl-Liebknecht-Str. 24/25, 14476 Potsdam, Germany
}

\date{\today}

\pubyear{2025}

\begin{document}

\label{firstpage}
\pagerange{\pageref{firstpage}--\pageref{lastpage}}
\maketitle

\begin{abstract}
The analogues of $\gamma$ Cassiopea are binary early type Be stars which are X-ray bright with hard thermal spectra. The nature of companions in these stars and mechanisms of their X-ray emission remain enigmatic. Among the proposed ideas  is  the presence of an accretion disc around a white dwarf (WD) companion to the Be star donor. We use radiative transfer models including reflection physics in order to calculate the synthetic spectra of such systems, and assume that the hottest plasma is thermal and is located in the accretion disc boundary layer. The models are used to analyse the {\it XMM-Newton} observations of the $\gamma$ Cas analogue $\zeta$ Tau (a.k.a. Ti\={a}ngu\={a}n). Comparisons with X-ray-emitting symbiotic systems, particularly $\delta$- and $\beta/\delta$-type systems, support the idea that the hard X-ray emission in $\zeta$ Tau is best explained by a WD accreting material expelled from the Be star. The plasma temperature and luminosity of the boundary layer associated with the accretion disc are used to estimate a mass accretion rate of $\dot{M}_\mathrm{acc} \approx 4\times 10^{-10}$ M$_\odot$~yr$^{-1}$, implying a nova recurrence time above 10$^{5}$ yr. Our analysis advances the understanding the production of hard X-ray emission in $\gamma$ Cas analogues, further supporting the idea of accreting WDs as companions of Be-stars in these systems.
\end{abstract}



\section{Introduction}\label{introduction}
\label{sec:intro}

$\gamma$ Cas analogues are persistently X-ray bright ($L_\mathrm{X,0.2-100~keV} \approx 10^{32}$ erg~s$^{-1}$) early Be-type binary stars named after their prototype $\gamma$ Cassiopeiae (a.k.a. HD~5394). Their X-ray spectra are thermal and are well described by a soft plasma component with temperature $kT \approx 0.1$ keV, and much harder components with temperatures well above 5 keV. Another prominent feature in X-ray spectra of all $\gamma$ Cas analogues is the 6.4 keV Fe fluorescent emission line \citep[see review in][]{Smith2016}. Currently, dozens of $\gamma$ Cas analogues are known \citep[e.g.,][]{Naze2025}.

The origin of the dominant hard X-ray emission from $\gamma$ Cas analogues remains a topic of ongoing debate. The most accepted explanations include ({\it i}) interactions between a hypothetical magnetic field of the Be star and its decretion disc \citep{Robinson2000, SmithRobinson1999} and ({\it ii}) accretion onto a compact object, such as a neutron star \citep[NS;][]{White1982,Postnov2017,Rauw2024} or a white dwarf \citep[WD; e.g.,][]{Murakami1986,Haberl1995,Kubo1998,Hamaguchi2016,Gies2023,Tsujimoto2018,Gunderson2025}. We note that \citet{Langer2020} proposed an alternative idea where the strong wind from a sdOB star companion can collide with the Be star disc powering X-ray emission in $\gamma$ Cas analogues. Nevertheless, \cite{Naze2022} rejected  this scenario on theoretical and empirical grounds. Understanding the origin of X-rays from $\gamma$ Cas analogues has important implications for revealing the evolutionary channels leading to the productions of Be-type stars as well as elucidating the origin and end-points of binaries containing compact objects in general.

Among the family of $\gamma$ Cas analogues, Ti\={a}ngu\={a}n (a.k.a. $\zeta$ Tau) has special diagnostic importance \citep{Gies2007}. The system is observed nearly edge on, which provides unparallel view on the binary components among $\gamma$ Cas analogues \citep[see][and references therein]{Quirrenbach1997,Tycner2004,Carciofi2009,Cochetti2019}.
$\zeta$ Tau consists of an 11~M$_\odot$ Be star and a low-mass companion with a mass of $\sim$0.8--1~M$_\odot$ in a 133-day orbit \citep{Ruz2009,Gies2007}. 
{\it Chandra} X-ray observations of $\zeta$ Tau were presented by \citet{Naze2022}, identifying this Be binary star as a member of the group of $\gamma$ Cas analogues. 
Although the {\it Chandra} observations of $\zeta$ Tau were significantly affected by pile-up, \citet{Naze2022} nonetheless detected highly attenuated emission from the hot plasma component. A definitive detection of this component, along with the 6.4 keV Fe emission line, was achieved in subsequent {\it XMM-Newton} observations, as presented by \citet{Naze2024}. In that study, the authors show that the observed column density can be reconciled with values estimated for the Be disk, however, the absorption variability estimated from the X-ray analysis does not correlate with the H$\alpha$ emission from the disc.

The best fitting models of the {\it XMM-Newton} spectra presented in \citet{Naze2024} include several optically-thin plasma emission components, with the flux of the softest temperature components associated with the Be star. The flux of the soft component was estimated to be consistent with the expected values for Be stars, with log$_{10}(L_\mathrm{X}/L_\mathrm{bol}) \gtrsim -8.6$. \citet{Naze2024} presented an extensive discussion trying to put into context the hard X-ray emission from $\zeta$ Tau with the presence of an accreting WD companion, although their results remained inconclusive.

Accreting WDs are found in a variety of binary systems \citep[see review by][]{Mukai2017}, with symbiotic systems relevant in the context of $\gamma$ Cas analogues \citep[see][]{Tsujimoto2018,Naze2024}. These systems are classified according to the $\alpha$, $\beta$, $\delta$, and $\beta/\delta$-type scheme, with each class exhibiting distinct X-ray spectral properties \citep[see][for details]{Murset1997, Luna2013}.  Particularly, $\delta$-type symbiotic systems are characterized by the presence of a heavily-absorbed ($N_\mathrm{H} \approx 10^{23}$--$10^{24}$ cm$^{-2}$) hot plasma ($kT \approx 5$--10 keV), along with the contribution from the 6.4 keV Fe fluorescent emission line \citep{Eze2014}. The $\beta/\delta$-type systems share these properties but also present a significant contribution in the soft X-ray regime \citep[$E < 1$ keV; e.g.,][]{Zhekov2019}.

The X-ray spectra of $\delta$- and $\beta/\delta$-type symbiotic systems are well-described by a multi-temperature plasma model with the {\it ad hoc} addition of a Gaussian profile to fit the 6.4 keV Fe emission line \citep[see][and references therein]{Lima2024}. The hard X-ray-emitting plasma in these systems is produced at the boundary layer \citep{Pringle1979,Patterson1985}, the region between the accretion disc and the surface of the accreting WD, while softer temperatures in $\beta/\delta$ symbiotics are attributed to winds and/or jets \citep[e.g.,][]{Lucy2020}. On the other hand, it has been established for decades that the 6.4 keV Fe emission originates from the reflection of X-ray photons by cooler, dense material in the accretion disc around the WD \citep[e.g.,][]{Ishida2009, Wheatley2006}. To consistently model the Fe K line in a physically motivated way, \citet{Toala2024} developed tools to incorporate reflection physics into the analysis of X-ray spectra from accreting WDs. Radiative-transfer simulations of X-ray photons passing through a dense medium, such as the accretion disc around the WD, are performed to consistently estimate the absorption and the reflection properties of the X-ray spectra. This approach aids in disentangling the various components of the X-ray emission from accreting WDs in symbiotic systems in a clearer way \citep[see][]{Toala2023,Toala2024_TCrB,VT2024}.

In this paper we present the analysis of the {\it XMM-Newton} observations of $\zeta$ Tau by means of state-of-the-art reflection models generated with radiative-transfer simulations. The paper is organised as follows. In Section~\ref{sec:obs} we briefly describe the observations and data analysis. We present radiative-transfer models of the X-ray emission through a disc in Section~\ref{sec:reflection}. The details of the spectral analysis and the results are presented in Section~\ref{sec:analysis}. The discussion is presented in Section~\ref{sec:discussion}. Finally, we present our summary and conclusions in Section~\ref{sec:summary}.

\section{Observations and data reduction}
\label{sec:obs}

$\zeta$ Tau has been observed by {\it XMM-Newton} using the three European Photon Imaging Cameras (EPIC), namely, pn, MOS1 and MOS2. Two sets of observations were carried out on 2023 March 20 and 2023 Oct 3 using the thick optical blocking filter and correspond to the Obs. ID. 0920020301 and 0920020401 (PI: Y. Naz\'{e}). The total observing times for the first epoch is 17 ks while that of the second epoch is 22 ks. The first observation campaign was performed with the Be in front of the companion ($\varphi=0.0$), and in the the second epoch the situation is inverted ($\varphi=0.5$).

The data were processed with the Science Analysis System \citep[{\sc sas}, version 20.0;][]{Gabriel2004} following the same procedure as that described by \citet{Naze2024}. The event files were created with the \texttt{epproc} and \texttt{emproc} tasks filtering the events with pattern 0--12 and 0--4 for the MOS and PN cameras, respectively. A spectrum per observation was extracted with the {\sc sas} task \texttt{eveselect}, using a circular region of 30~arcsec centred on $\zeta$ Tau. A background spectrum was extracted from the vicinity of $\zeta$ Tau similarly to that illustrated in figure 1 of \citet{Naze2024}. Finally, the response matrices were created using the {\sc sas} tasks \texttt{rmfgen} and \texttt{arfgen}.
We use only the EPIC-pn spectra in this work given the superior effective area of this camera compared with the EPIC MOS instruments.

After data processing, the first and second epoch of observations resulted in EPIC-pn count rates of 637$\pm$9 and 795$\pm$8 counts~ks$^{-1}$ for the 0.3--10.0 keV energy range. That is, $\approx$4840 and $\approx$9300 counts for net exposure times of 7.6 and 11.7 ks, respectively.

\begin{figure}
\begin{center}
\includegraphics[width=0.95\linewidth]{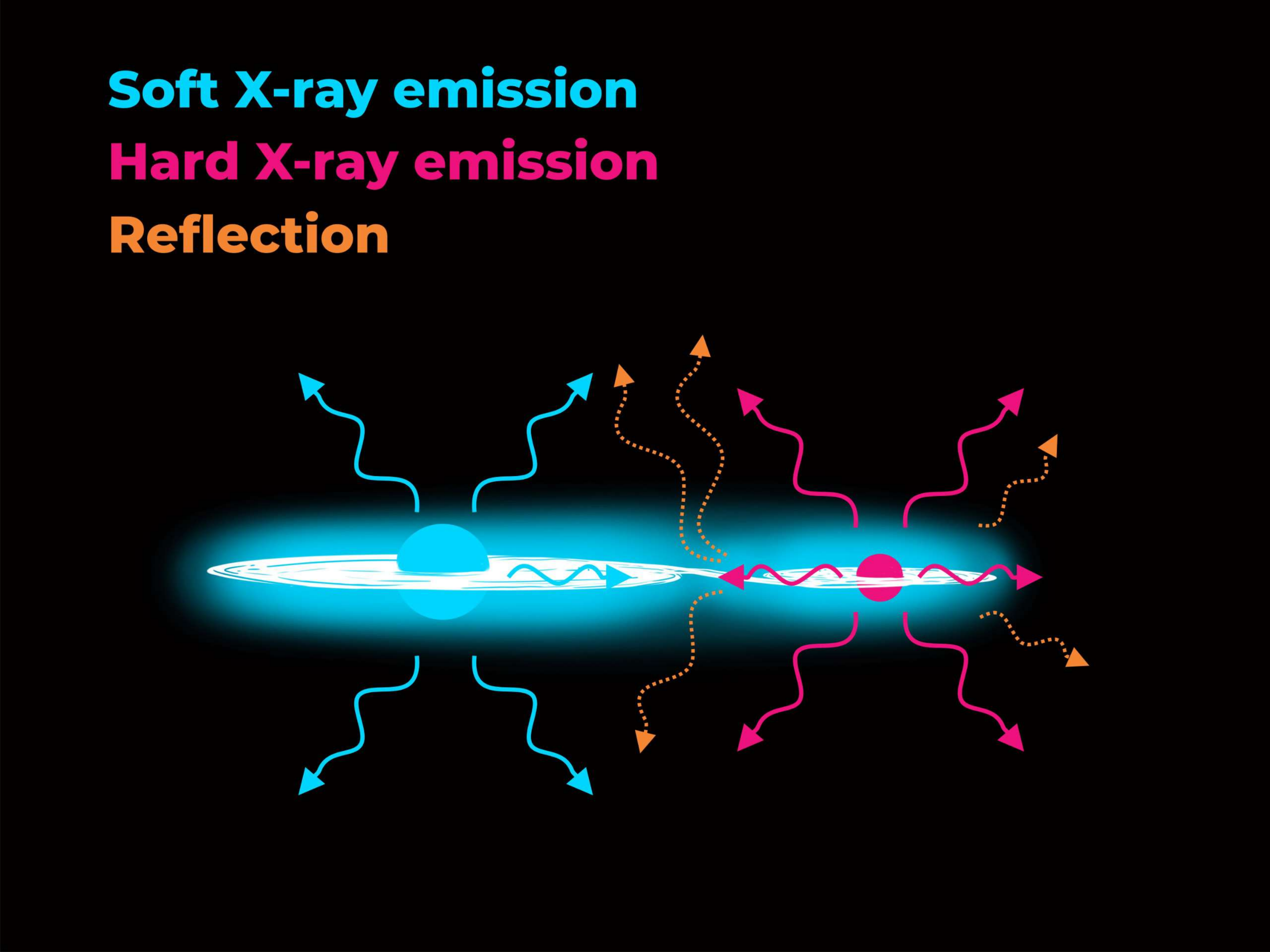}
\end{center}
\caption{A schematic representation of the decretion/accretion discs in $\zeta$ Tau. The contributions from the soft and hard X-ray emissions are shown, as well as the production of reflection. The Be star is the source of soft X-ray emission, while the boundary layer around accreting WD produced the hard X-ray emission.}
\label{fig:sketch}
\end{figure}

\section{Synthetic X-ray spectra of accretion discs including reflection}
\label{sec:reflection}

\begin{figure}
\begin{center}
\includegraphics[width=\linewidth]{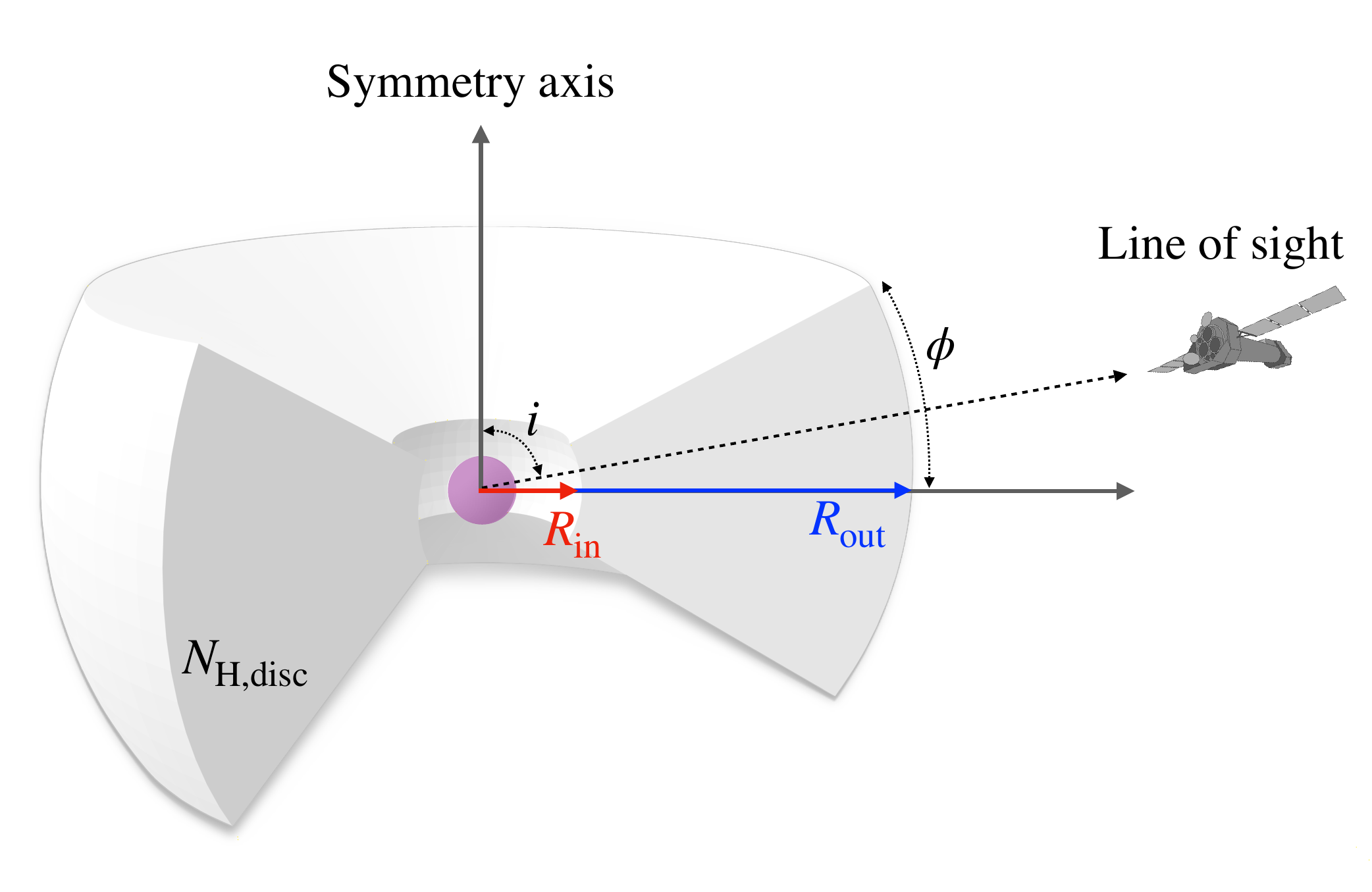}
\end{center}
\caption{A schematic representation of a flared disc around the accreting component in $\zeta$ Tau. The angle $i$ is the inclination of the symmetry axis with respect to the line of sight, while $\phi$ is the opening angle of the flared disc. The boundary layer corresponds to the region between the surface of compact object and $R_\mathrm{in}$.}
\label{fig:disc}
\end{figure}

Motivated by the apparent similarities between the X-ray spectra of $\zeta$ Tau and those of $\beta/\delta$-type symbiotic systems, we make an ansatz that the hard X-ray emission has a reflection component, an idea that is supported by the presence of the 6.4 keV Fe fluorescent line in X-ray spectra. We note that similar ideas were explored by \citet{Tsujimoto2018} for the X-ray analysis of $\gamma$ Cas and HD~110432. Here, we propose that the low-mass component in $\zeta$ Tau accretes material from material expelled from the Be star (winds and the decretion disc), forming its own accretion disc. This idea follows from numerical simulations of Be/X-ray binaries, where a NS is accreting from its Be star companion \citep[see, e.g.,][and references therein]{Franchini2021,MacLeod2023,Martin2014,Martin2024,Okazaki2002,Rast2025}. Those simulations demonstrate the formation of an accretion disc around the compact object, creating an accretion/decretion disc system. A sketch of the proposed topology of $\zeta$ Tau (and $\gamma$ Cas analogues in general) is presented in Fig.~\ref{fig:sketch}. Note however, that the mass of the compact object in $\zeta$ Tau suggests that it is a WD rather than a NS.

We explore synthetic X-ray spectra and their reflection properties due to the presence of an accretion disc by means of the radiative-transfer code {\sc skirt} \citep[version 9;][]{Camp2020}. {\sc skirt} accounts for the effects of Compton scattering on free electrons, photoabsorption, and fluorescence by cold atomic gas, scattering on bound electrons and extinction by dust \citep[see][]{vanderMeulen2023}.

The radiative transfer simulations are performed adopting a flared accretion disc geometry (see Fig.~\ref{fig:disc}). The disc is defined mainly by four parameters: the inner ($R_\mathrm{in}$) and outer radii ($R_\mathrm{out}$), the opening angle ($\phi$), and an averaged hydrogen column density ($N_\mathrm{H,disc}$).

In principle, the source of hard X-ray photons in an accretion disc is the region delimited by the inner radius of the accretion disc and the surface of the accreting compact object, known as the boundary layer. In our simulations we approximate this region by defining our source of hard X-ray photons as a point source at the origin of the density grid. We assume that the X-ray radiation is produced by thermal, collisionally-ionised plasma. To describe its spectrum, we use the \texttt{apec} model \citep{Smith2001}. The plasma temperature was set to 9 keV \citep[see][and Section~\ref{sec:analysis}]{Naze2024} with abundances fixed to the solar values of \citet{Lodders2009}.

\begin{figure}
\begin{center}
\includegraphics[width=0.95\linewidth]{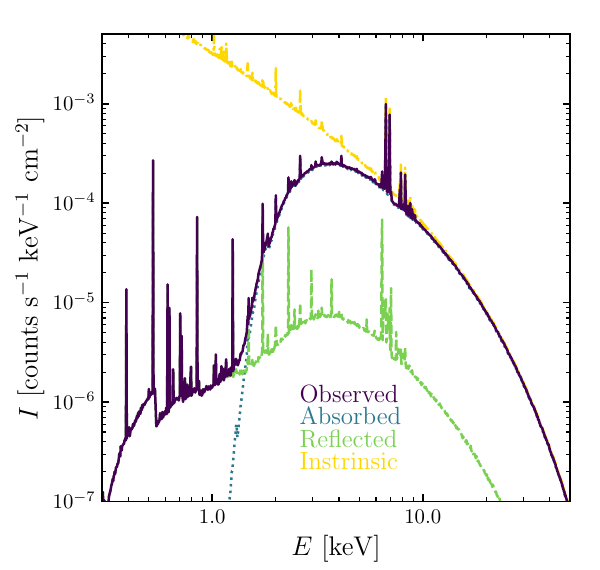}
\end{center}
\caption{Synthetic X-ray spectra obtained from a radiative-transfer simulation with {\sc skirt} adopting $N_\mathrm{H,disc}=2\times10^{23}$~cm$^{-2}$, inner and outer radii of $R_\mathrm{in}=0.1$ R$_\odot$ and $R_\mathrm{out} = 0.3$ AU, input plasma temperature of 9 keV for the boundary layer, computed for an inclination angle of $i=85^\circ$. The observed (solid), absorbed (dotted), reflected (dashed), and intrinsic (dash-dotted) spectra are illustrated.}
\label{fig:skirt}
\end{figure}

Fig.~\ref{fig:skirt} presents an example of a synthetic X-ray spectrum in the 0.3--50 keV energy range for a disc with $N_\mathrm{H,disc}=2\times10^{23}$~cm$^{-2}$ under viewing angle of $i=85^{\circ}$. The figure describes the components of the emergent spectrum: the intrinsic spectrum at the boundary layer, the spectrum attenuated by the material in the disc, and the reflected component also produced by the disc. It is noteworthy that the 6.4 keV Fe emission line is naturally produced by reflection in the disc, while the 6.7 and 6.97 He- and H-like Fe lines are produced by the thermal plasma in the boundary layer \citep[see][]{Ishida2009,Eze2014}. This figure also shows that the contribution from the Compton reflection continuum (dashed line in Fig.~\ref{fig:skirt}) is not dominant in the observed spectrum. An inverted situation is predicted by radiative-transfer simulations of the $\beta/\delta$ symbiotic system R Aqr for energies above 12 keV \citep[][]{VT2024}.

As a next step we compute a grid of models for different values of $R_\mathrm{out}$ and $N_\mathrm{H,disc}$. The values of $R_\mathrm{out}$ are between 0.1 and 0.6 AU\footnote{The Roche lobe radius is $\approx$0.68 AU for $\zeta$ Tau.}, while for $N_\mathrm{H,disc}$ we adopt values between $0.8\times10^{23}$ and $1\times10^{24}$. The inner radius of the disc is fixed to a reasonable value of $R_\mathrm{in}= 0.1$ R$_\odot$ ($\approx4.7\times10^{-4}$ AU) for an accreting WD and the inclination is $i=85^\circ$, that is, close to an edge-on view. The opening angle of the flared disk is fixed to $\phi=20^{\circ}$, however, we note that other values ($\phi=10,15,25$ and 30$^{\circ}$) produce similar results given that the disc is seen almost edge-on.

\begin{figure}
\begin{center}
\includegraphics[width=0.95\linewidth]{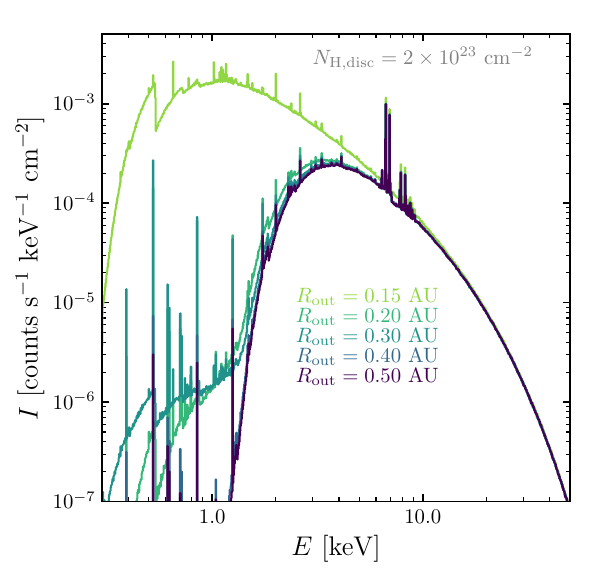}\\
\includegraphics[width=0.95\linewidth]{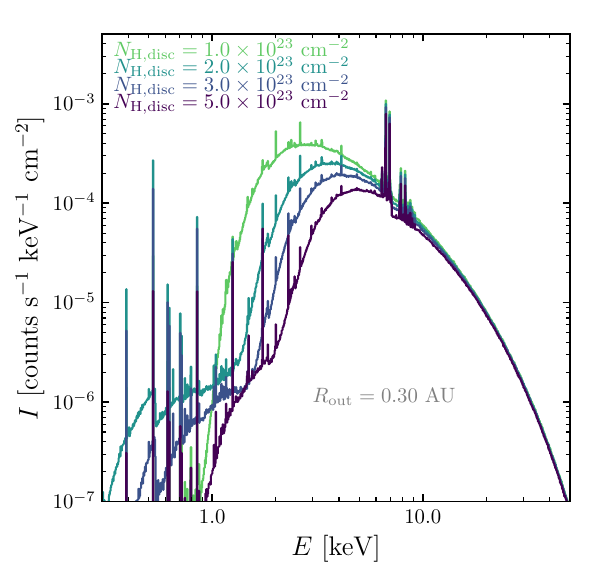}
\end{center}
\caption{Synthetic X-ray spectra obtained from radiative transfer {\sc skirt} simulations. {\bf Top:} A family of models with same disc column density ($N_\mathrm{H,disc}=2\times10^{23}$ cm$^{-2}$) varying the outer radius ($R_\mathrm{out}$) of the accretion disc. {\bf Bottom:} A second family of models adopting the same outer radius of $R_\mathrm{out}=0.3$ AU but different $N_\mathrm{H,disc}$.}
\label{fig:comparison}
\end{figure}

Examples of model grid X-ray spectra are shown in Fig.~\ref{fig:comparison}. 
It can be noted that the combination of absorption and reflection effects are almost indistinguishable for discs with $R_\mathrm{out} \geq 0.2$ AU, while larger values of $N_\mathrm{H,disc}$ displace the maximum of spectral energy distribution towards higher energies.

\section{Spectral analysis}
\label{sec:analysis}

\begin{figure}
\begin{center}
\includegraphics[width=0.95\linewidth]{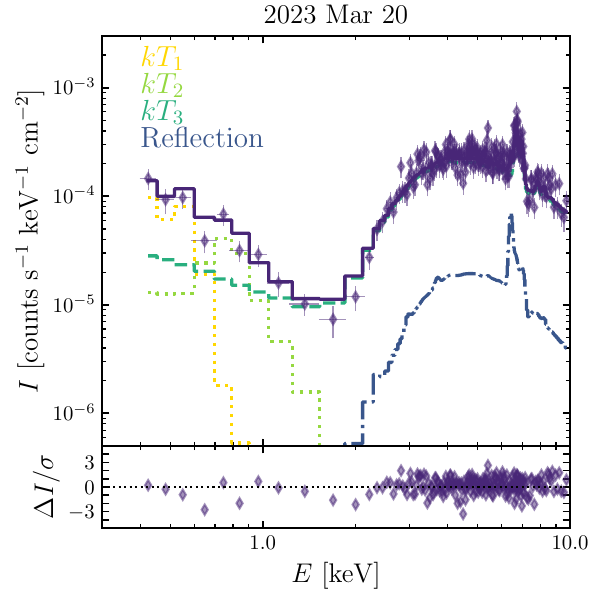}\\
\includegraphics[width=0.95\linewidth]{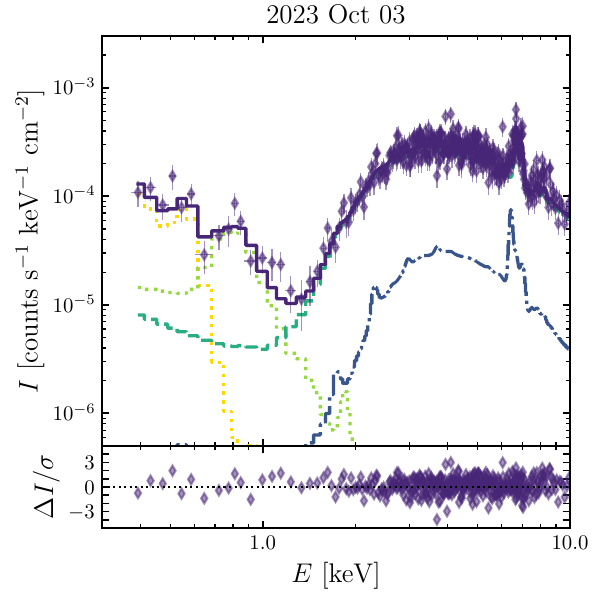}
\end{center}
\caption{Background-subtracted {\it XMM-Newton} EPIC-pn spectra of $\zeta$ Tau (diamonds) in the 0.3--10 keV range. The top and bottom panels show the results from observations obtained on 2023 Mar 20 and 2023 Oct 03 that correspond to orbital phases of $\varphi=0.0$ and $\varphi=0.5$, respectively. In both epochs the best-fit model is presented with a solid line. Different components are shown with different line styles. See model parameters in Table~\ref{tab:analysis}.}
\label{fig:spectra}
\end{figure}

The two {\it XMM-Newton} EPIC-pn spectra of $\zeta$ Tau are almost identical, but there is a small shift between the two spectra in the 1--3 keV band \citep[see Fig.~\ref{fig:spectra} and fig.~6 in][]{Naze2024}. Such difference might be attributed to a lower column density during the second epoch of observation (see, e.g, Fig.~\ref{fig:comparison}).
Other subtle differences might be attributed to the different intensities of some of the emission lines. For example, the $\lesssim$0.8 keV feature which might be a blending of the O~{\sc viii} and Ne~{\sc ix} emission lines.

The analysis of the EPIC-pn spectra has been performed by means of the X-Ray Spectral Fitting Package \citep[{\sc xspec}, version 12.12.1;][]{Arnaud1996}. The thermal plasma emission is modelled using \texttt{apec}
while the absorption is included with a Tuebingen-Boulder ISM absorption model \citep[\texttt{tbabs};][]{Wilms2000}. We start from simple models, and increase model complexity depending on the quality of the fit, which is assessed by evaluating the reduced $\chi^{2}$ statistics ($\chi^{2}_\mathrm{DoF}$).

In order to include the reflection component into the spectral analysis, our radiative-transfer model grid produced with {\sc skirt} is converted into additive one-parameter tables (in fits format) using the {\sc heasoft}\footnote{\url{https://heasarc.gsfc.nasa.gov/docs/software/heasoft/}} task \texttt{ftflx2tab}. For each of the {\sc skirt} models, only the reflected component is included in {\sc xspec} (e.g., dashed line in Fig.~\ref{fig:skirt}). In the following, this is referred as \texttt{reflection} components.

Motivated by our experience with $\delta$- and $\beta/\delta$-type  symbiotic systems, we start with a spectral model that includes an absorbed plasma plus the reflection component, 
\begin{equation}
   \texttt{tbabs*CF*(apec+reflection)},
\end{equation}
\noindent where CF represents a covering factor that describes possible non-uniform distribution of material in the disc. This model fits well the hard X-ray spectral range ($E>1$ keV), but fails to reproduce the soft X-ray emission resulting in a bad quality fit ($\chi^2_\mathrm{DoF} > 2$). Thus, we add another spectral model component that describes soft thermal emission with an independent absorbing column density corresponding to the ISM absorption
\begin{equation}
   \texttt{tbabs}_\texttt{ISM} \texttt{*apec}_\texttt{1} \texttt{ + tbabs}_\texttt{2} \texttt{*CF*(apec}_\texttt{2} \texttt{+reflection)}.
\end{equation}
\noindent However, similarly to obtained by the spectral analysis presented by \citet{Naze2024}, the soft X-ray emission seem to be best described by a multi-temperature component. Consequently, we also attempted models of the form
\begin{equation}
\begin{split}
   \texttt{tbabs}_\texttt{ISM} & \texttt{*}\texttt{(apec}_\texttt{1}\texttt{+}\texttt{apec}_\texttt{2} \texttt{)} \\ 
   &\texttt{ + tbabs}_\texttt{2} \texttt{*CF*(apec}_\texttt{3} \texttt{+reflection)}.
\end{split}
\label{eq:3}
\end{equation}
\noindent The hydrogen column density provided by the ISM was fixed to $N_\mathrm{H,ISM}=2.7\times10^{20}$~cm$^{-2}$ \citep[][]{Diplas1994}.
The model configuration defined by Eq.~(\ref{eq:3}) resulted in good quality fits for both epochs ($\chi^2_\mathrm{DoF} \approx 1$; see Table~\ref{tab:analysis}), they are plotted alongside the observed EPIC-pn spectra in Fig.~\ref{fig:spectra}. We remark that the best-fits were achieved by reflection models with $R_\mathrm{out}=0.3$ AU.

\begin{table}
\begin{center}
\caption{Parameters of the best-fit models obtained for the two {\it XMM-Newton} observations of $\zeta$ Tau. The observed ($f_\mathrm{X}$) and intrinsic ($F_\mathrm{X}$) fluxes are computed for the 0.3--12.0 keV band. The X-ray luminosity ($L_\mathrm{X}$) is calculated adopting a distance of 136 pc. $N_\mathrm{H,ISM}$ is fixed during the analysis.}
\label{tab:analysis}
\setlength{\tabcolsep}{0.6\tabcolsep}  
\begin{tabular}{llcc}
\hline
Parameter & Units & 2023 Mar 20 & 2023 Oct 03\\
\hline
$\chi^{2}_\mathrm{DoF}$ &  & 213.96/214=1.00 & 406.33/393 = 1.03 \\
\hline
Soft Emission \\
$N_\mathrm{H,ISM}$  & [10$^{22}$~cm$^{-2}$]    & 2.7$\times10^{-2}$              & 2.7$\times10^{-2}$      \\
$kT_1$           & [keV]                    & 0.08$\pm$0.03                   & $0.09\pm$0.02\\
$A_1$            & [cm$^{-5}$]              & (3.8$\pm$0.7)$\times$10$^{-4}$  & (2.4$\pm$0.4)$\times$10$^{-4}$ \\
$f_\mathrm{X1}$  & [erg~cm$^{-2}$~s$^{-1}$] & (5.0$\pm$0.9)$\times$10$^{-14}$ & (4.0$\pm$0.6)$\times$10$^{-14}$\\
$F_\mathrm{X1}$  & [erg~cm$^{-2}$~s$^{-1}$] & (7.8$\pm$1.3)$\times$10$^{-14}$ & (6.1$\pm$0.9)$\times$10$^{-14}$\\
$L_\mathrm{X1}$  & [erg~s$^{-1}$]           & (1.7$\pm$0.3)$\times$10$^{29}$  & (1.3$\pm$0.2)$\times$10$^{29}$ \\
$kT_2$           & [keV]                    & 0.50$\pm$0.18                   & 0.66$\pm$0.07\\
$A_2$            & [cm$^{-5}$]              & (9.6$\pm$3.3)$\times$10$^{-2}$  & (1.2$\pm$0.3)$\times$10$^{-5}$ \\
$f_\mathrm{X2}$  & [erg~cm$^{-2}$~s$^{-1}$] & (2.0$\pm$0.7)$\times$10$^{-14}$ & (2.8$\pm$0.6)$\times$10$^{-14}$\\
$F_\mathrm{X2}$  & [erg~cm$^{-2}$~s$^{-1}$] & (2.3$\pm$0.8)$\times$10$^{-14}$ & (3.1$\pm$0.7)$\times$10$^{-14}$\\
$L_\mathrm{X2}$  & [erg~s$^{-1}$]           & (5.0$\pm$1.8)$\times$10$^{28}$  & (6.8$\pm$1.5)$\times$10$^{28}$ \\
\hline
Hard Component \\
$N_\mathrm{H2}$  & [10$^{22}$~cm$^{-2}$]    & 14.0$\pm$0.4                    & 7.6$\pm$0.2\\
CF               &                          & 0.997$\pm$0.001                 & 0.998$\pm$0.002\\
$kT_3$           & [keV]                    & 9.0$\pm$0.5                     & 9.5$\pm$0.4\\
$A_3$            & [cm$^{-5}$]              & (1.5$\pm$0.1)$\times$10$^{-2}$  & (1.3$\pm$0.1)$\times$10$^{-2}$\\
$f_\mathrm{X3}$  & [erg~cm$^{-2}$~s$^{-1}$] & (1.1$\pm$0.1)$\times$10$^{-11}$ & (1.2$\pm$0.1)$\times$10$^{-11}$\\
$F_\mathrm{X3}$  & [erg~cm$^{-2}$~s$^{-1}$] & (3.1$\pm$0.1)$\times$10$^{-11}$ & (2.6$\pm$0.1)$\times$10$^{-11}$\\
$L_\mathrm{X3}$  & [erg~s$^{-1}$]           & (6.7$\pm$0.3)$\times$10$^{31}$  & (5.7$\pm$0.1)$\times$10$^{31}$ \\
Reflection \\
$A_\mathrm{ref}$            & [cm$^{-5}$]   & 0.13$\pm$0.04                   & 0.16$\pm$0.04 \\
$f_\mathrm{ref}$  & [erg~cm$^{-2}$~s$^{-1}$]& (9.7$\pm$0.3)$\times$10$^{-13}$ & (1.2$\pm$0.4)$\times$10$^{-12}$\\
$F_\mathrm{ref}$  & [erg~cm$^{-2}$~s$^{-1}$]& (1.7$\pm$0.6)$\times$10$^{-12}$ & (1.9$\pm$0.5)$\times$10$^{-12}$\\
$L_\mathrm{ref}$  & [erg~s$^{-1}$]          & (3.7$\pm$1.3)$\times$10$^{30}$  & (4.1$\pm$1.2)$\times$10$^{30}$ \\
\hline
$f_\mathrm{X,TOT}$& [erg~cm$^{-2}$~s$^{-1}$]& (1.2$\pm$0.1)$\times$10$^{-11}$ & (1.3$\pm$0.1)$\times$10$^{-11}$\\
$F_\mathrm{X,TOT}$& [erg~cm$^{-2}$~s$^{-1}$]& (3.2$\pm$0.2)$\times$10$^{-11}$ & (2.8$\pm$0.2)$\times$10$^{-11}$\\
$L_\mathrm{X,TOT}$& [erg~s$^{-1}$]          & (7.1$\pm$0.5)$\times$10$^{31}$  & (6.2$\pm$0.3)$\times$10$^{31}$ \\
\hline
\end{tabular}
\end{center}
\end{table}

Table~\ref{tab:analysis} lists the observed flux ($f_\mathrm{X}$) as well as the flux corrected by absorption ($F_\mathrm{X}$) integrated for the 0.3--10.0 keV energy range. The corresponding X-ray luminosities ($L_\mathrm{X}$) have been computed adopting a distance of $d=136$ pc. In addition, Table~\ref{tab:analysis} also shows the contribution from each model component to the observed and intrinsic fluxes as well as their luminosities. The normalization parameter $A$ for each component is also listed\footnote{The normalization parameter is defined as $A \approx 10^{-14}\int n_\mathrm{e}^2 dV/4 \pi d^2$, where $n_\mathrm{e}$ is the electron number density and $V$ is the X-ray-emitting volume.}.

\begin{figure}
\begin{center}
\includegraphics[width=0.95\linewidth]{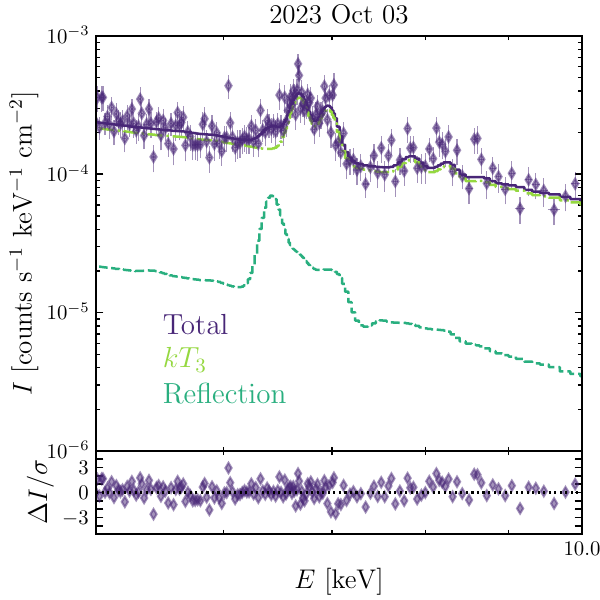}
\end{center}
\caption{Comparison between the EPIC-pn spectra of the 2023 Oct 3 epoch and the best-fit model in the 5.0--10.0 keV range to highlight the contribution from the reflection component to the 6.4 keV Fe emission line and that of the hot plasma producing the 6.7 and 6.97 keV He- and H-like Fe lines.}
\label{fig:iron}
\end{figure}

The spectral analysis presented here suggests that, within the error margins, there are no significant changes in the total X-ray fluxes and luminosities between the two epochs of observations. The temperature of the boundary layer remains relatively constant ($kT_3 \approx$ 9 keV) in both epochs. On the other hand, the absorbing column density of the hottest plasma ($N_\mathrm{H2}$) located at the boundary layer, is indeed larger in the first epoch \citep[see also Table 1 in][]{Naze2024}. The variations in $N_\mathrm{H2}$ between the two epochs are likely due to changes in the density structure of the accretion disc or to the fact that, in the first epoch, the compact companion is located behind the Be star.

The plasma temperatures of the two components fitting the soft X-ray emission resulted in $kT_1 = 0.08$ keV and $kT_2 \approx 0.5$--0.7 keV, almost the same values as those reported by \citet{Naze2024}. The intrinsic X-ray luminosity of the soft X-ray emission resulted in $L_\mathrm{soft}=L_\mathrm{X1} + L_\mathrm{X2} \approx 2\times10^{29}$~erg~s$^{-1}$ for both epochs. If this component is associated with the Be star in $\zeta$ Tau, we can adopt $L_\mathrm{bol}$=5620~L$_\odot$ \citep[see][]{Naze2022} and estimate log$_{10}(L_\mathrm{X}/L_\mathrm{bol})=-8.04$, which agrees with calculations performed in \citet{Naze2024} and is similar to other early type Be stars \citep[][]{Naze2018,Naze2023}.

A consistent inclusion of the reflection physics through radiative-transfer simulations avoids the arbitrary inclusion of a Gaussian profile to model the 6.4 keV Fe fluorescent line. Our analysis of $\zeta$ Tau shows that the reflection component represents about 5--6 per cent of the flux of the hard X-ray emission. This is fully enough to explain the presence of the 6.4 keV Fe fluorescent line as shown in Fig.~\ref{fig:iron}. 

Finally, we note that our analysis did not required the correction of the EPIC spectra for charge transfer efficiency calibration effects, as described in \citet{Naze2024}. The quality of our models in the energy range where the 6.4, 6.7, and 6.97 keV Fe emission lines appear to have acceptable residual values (see bottom panel in Fig.~\ref{fig:iron}).

\section{Discussion}
\label{sec:discussion}

We propose a model that attributes the bulk of hard X-ray emission in the binary Be-type star $\zeta$ Tau to the boundary layer of the accretion disc around the WD companion. 
The hypothesis is tested by fitting the observed multi-epoch  {\em XMM-Newton} X-ray spectra 
with our new  models computed with the {\sc skirt} radiative transfer code which is used to describe the reflection of X-rays from the accretion disc. In the spectral modelling, the reflection component is combined with the emission from thermal plasma located at the boundary layer.

The good spectral fits we have achieved further support the idea of  $\gamma$ Cas analogues being binary systems consisting of an  early type Be star and an accreting WDs. 
Furthermore, our analysis has demonstrated that the X-ray luminosity associated with the boundary layer around the accreting companion in $\zeta$ Tau, $L_\mathrm{X3}$, is consistent with the peak of the luminosity distribution for X-ray-emitting symbiotic systems \citep[see fig.~5 in][]{Guerrero2024}.

While the idea of $\gamma$ Cas analogues being binary stars hosting compact objects have been previously discussed in the literature, but the quantitative analyses, such as presented in this paper, have been very scarce. \citet{Tsujimoto2018} used similar approach to analyse X-ray spectra of two other $\gamma$ Cas-type stars. They considered two cases:  of a non-magnetic and a magnetic WDs, and concluded that both explain the observations. It is interesting to note that the analysis of the X-ray variability in $\gamma$ Cas led \citet{Hamaguchi2016} to favour a magnetized WD model. \citet{Rauw2024} computed the emissivity of the Fe K line in the X-ray spectra of $\gamma$ Cas  to test the quantitative theoretical predictions for a propelling NS and demonstrated  discrepancies between the model predictions and observed Fe K line. 

The analysis presented in this paper is motivated by the apparent similarities between X-ray spectra of $\zeta$ Tau and those of accreting WDs in symbiotic systems, particularly 
$\delta$- and $\beta/\delta$-type objects.  
Our results agree with previous works which describe the X-ray 
spectra of $\gamma$ Cas analogues as consisting of two spectral components, a softer and a harder one, which is similar to symbiotic systems. 

The soft component in spectra of symbiotics is typically attributed to plasma heated by jets and/or disc winds \citep[see, for example,][]{Lucy2020, Toala2022, Toala2024_TCrB}. In $\gamma$ Cas analogues which have early type Be star donors, the soft spectra are naturally explained by X-ray emission from the donor's stellar wind.

The hard spectral component in the symbiotic systems of type $\delta$ and $\beta/\delta$
originates from plasma with a temperature of $\sim100$~MK, located in the
boundary layer of an accretion disc. The radiation emerging from this region is heavily absorbed and partly reflected by the disc. The reflection component is responsible for Fe\,K line at 6.4\,keV. We propose that similar mechanism operates in $\gamma$ Cas analogues. In Fig.~\ref{fig:iron} we compare a segment of the EPIC-pn spectrum with our best-fit model. The comparison shows that our model successfully reproduces the 6.4 keV Fe fluorescent line,  without including an {\em ad hoc} Gaussian component, as still done in the analysis of $\gamma$ Cas analogues in the literature \citep[e.g.,][]{Naze2024,Gunderson2025}. The other Fe emission lines at 6.7 and 6.97 keV are also naturally explained by the hot plasma in the boundary layer.

According to analytical predictions of accreting WDs \citep{Pringle1979,Patterson1985}, the temperature of the boundary layer can vary between $0.1$\,MK to $100$\,MK, depending on the mass accretion rate $\dot{M}_\mathrm{acc}$. Higher $\dot{M}_\mathrm{acc}$ increases the density of the boundary layer making the cooling more efficient and, consequently, reducing the temperature.  Conversely, lower $\dot{M}_\mathrm{acc}$  results in higher plasma temperatures.

We estimate $\dot{M}_\mathrm{acc}$ in $\zeta$ Tau by two different methods. Firstly, according to \citet{Patterson1985}, the temperature of the boundary layer scales as  
\begin{equation}
    kT \approx 1.3 \left(\frac{M_\mathrm{WD}}{0.7~\mathrm{M}_\odot}\right)^{3.6} \left( \frac{10^{16}~\mathrm{g}~\mathrm{s}^{-1}}{\dot{M}_\mathrm{acc}} \right),
    \label{eq:acc1}
\end{equation}
\noindent where $M_\mathrm{WD}$ denotes the mass of the accreting WD. Adopting $M_\mathrm{WD}=0.8$ M$_\odot$ \citep[see][]{Ruz2009}, Eq.~(\ref{eq:acc1}) suggests $\dot{M}_\mathrm{acc}\approx6\times10^{-10}$~M$_\odot$~yr$^{-1}$.

Secondly, the accretion powered luminosity $L_\mathrm{acc}$  can be expressed as \citep{Shakura1973,Frank2002}
\begin{equation}
L_\mathrm{acc} \approx \frac{G M_\mathrm{WD}}{R_\mathrm{WD}} \dot{M}_\mathrm{acc},
\label{eq:acc2}
\end{equation}
\noindent were $R_\mathrm{WD}$ is the radius of the WD and $G$ is the gravitational constant. According to the discussion presented by \citet{Toala2024_TCrB}, the X-ray luminosity of the boundary layer in symbiotic systems (in our case $L_\mathrm{X3}$) is related to $L_\mathrm{acc}$ through an efficiency factor $\mu$ as $L_\mathrm{acc}=\mu L_\mathrm{X3}$. Adopting a typical value of $R_\mathrm{WD}=0.01$ R$_\odot$ and an efficiency value of 
$\mu = 20$--$100$ \citep[see discussion in][]{Toala2024_TCrB} we estimate $\dot{M}_\mathrm{acc}\approx (2$--$7)\times10^{-10}M_\odot$~yr$^{-1}$. Both estimates of the mass accretion rate are consistent  and are compatible with the mass-loss rates of stellar winds in early type B-stars \citep{Oskinova2011}. Interestingly, \citet{Tsujimoto2018} estimates mass accretion rate in $\gamma$ Cas and HD~110432 systems as $\lesssim 2\times10^{-10}$~M$_\odot$~yr$^{-1}$, very similar to that estimated here for $\zeta$ Tau.

Adopting a conservative value of $\dot{M}_\mathrm{acc} = 4 \times 10^{-10}$~M$_\odot$~yr$^{-1}$, we can estimate the recurrence of  nova-like events in $\zeta$ Tau. According to figure 2 of \citet{Chomiuk2021}, an accreting WD with a mass of $M_\mathrm{WD}=0.8$ M$_\odot$ has a recurrence time period of $10^{5\,...\,6}$~yr, which explains why novae events have not yet been detected from $\gamma$ Cas analogues.

It is important to remark here that we are not suggesting $\zeta$ Tau and other $\gamma$ Cas analogues to be symbiotic systems\footnote{ Symbiotic stars are binary systems where a WD accretes material from a late-type star and produces any kind of electromagnetic signature \citep[see review by][]{Merc2025}.}. Instead, we suggest that the physics behind the production of heavily-absorbed X-ray emission is the same scenario as that of $\delta$ and $\beta/\delta$ symbiotic binaries: the presence of an accretion disc and a boundary layer around a WD companion. The idea of similar X-ray emission mechanisms (accretion onto a WD) in $\gamma$ Cas analogues and symbiotic binaries is further corroborated by a close physical analogy between $\alpha$-type symbiotics and supersoft Be+WD systems where the later accretes stellar wind directly and X-rays are powered by thermonuclear burning on the WD surface. The $\alpha$-type nova symbiotic systems and Be$+$WD systems have similar X-ray luminosities and X-ray spectra. This strongly suggest that the key difference among these classes is the donor type. While it is a late type star in symbiotics, it is an early type stars in Be$+$WD binaries \citep{Coe2020, Kahabka2006, Kennea2021, Li2012, Sturm2012, Murset1997, Luna2013, Orio2007}.

\citet{Zhu2023} proposed that Be+WD systems may arise from the evolution of binaries initially composed of a Be star and a subdwarf OB companion. Such systems are known in the Galaxy, and the modern population synthesis calculations suggest that the population of such stars is quite sizeable   \citep{Wang2023,Yungelson2024, Hovis-Afflerbach2025}. While only a few Be+WD systems have been discovered  so far \citep{Kennea2021},  binary evolution models predict that they should be  common -- possibly even more so than Be+NS systems -- with a WD mass distribution peaking between 0.8 and 1.0 M$_\odot$ \citep[see][]{Raguzova2001}. These predictions are consistent with the mass estimated for the compact object in $\zeta$~Tau, and secondary stars in $\gamma$ Cas analogues in general. Hence,  the method applied in the current work serves as a potential new  X-ray diagnostics tool for the  identification of the Be+WD systems, and solving the puzzle $\gamma$ Cas analogues.   Moreover, Be+WD systems are of particular interest as they may serve as progenitors of black hole+WD systems, red nova events, and double-degenerate WD binaries \citep{ShaoLi2021, Zhu2023}.

\section{Summary and Conclusions}
\label{sec:summary}

We suggest that the binary star $\zeta$ Tau (a.k.a. Ti\={a}ngu\={a}n) hosts a WD that accretes material from expelled material (the wind and decretion disc) of the Be-star donor, naturally producing an accretion disc. We apply the radiative-transfer code {\sc skirt} to compute a grid of spectral models which are then compared to the X-ray spectra of $\zeta$ Tau observed by the {\em XMM-Newton} telescope. We show that the models reproduce the observed spectra, naturally including the fluorescent of Fe K emission line at 6.4\,keV. Our main findings are:
\begin{itemize}
 
    \item The soft component in the X-ray spectra of $\zeta$ Tau can be attributed to the Be star wind emission as previously suggested. 

    \item 
    The EPIC spectra of $\zeta$ Tau closely resemble those of $\delta$ and $\beta/\delta$-type symbiotic systems. Applying spectral models assuming that strong absorption and reflection arise from the accretion disc around the WD component allows to consistently reproduce the Fe K emission line in X-ray spectra.

    \item The temperature of the plasma at the boundary layer, the region between the inner radius of the accretion disc and the surface of the accretor, resulted in $kT_3 \approx 9$  keV and its X-ray luminosity is $L_\mathrm{X3}\approx 6\times10^{31}$~erg~s$^{-1}$. These values are  consistent with estimations of accreting WDs in symbiotic systems. 

    \item The  mass accretion rate needed to explain the temperature of the boundary layer and its luminosity is $\dot{M}_\mathrm{acc} \approx 4 \times10^{-10}$~M$_\odot$~yr$^{-1}$. According to WD accretion models, this accretion rate suggests nova recurrence times 
    $> 10^{5}$~yr.

\end{itemize}

The analysis presented here sheds new light on the nature of the $\gamma$ Cas analogues. By refining the modelling of hard X-ray emission using reflection physics, our findings increasingly suggest that WDs are the likely accretors in these systems. Notably, the spectra of $\gamma$ Cas analogues with accreting WDs (including $\zeta$ Tau) align with the established classification used for symbiotic systems \citep[$\alpha, \beta, \delta$, and $\beta/\delta$;][]{Murset1997,Luna2013}. This unified framework underscores the critical role of accreting WDs across both types of systems, 
facilitating a more integrated approach to their classification and study.

\section*{Acknowledgements} 

The authors thank the referee, Sean Gunderson, for comments and suggestions that improve our manuscript. The authors thank J.~C.~\'{A}lvarez from X.i.k. studio for providing the cartoon used in Figure 1 of this paper. J.A.T. and D.A.V.T. acknowledges support from the UNAM PAPIIT project IN102324. D.A.V.T. thanks Secretar\'{i}a de Ciencia, Humanidades, Tecnolog\'{i}a e Innovaci\'{o}n (SECHTI, Mexico) for a student grant. Based on observations obtained with {\it XMM-Newton}, an ESA science mission with instruments and contributions directly funded by ESA Member States and NASA. This work has made extensive use of NASA's Astrophysics Data System. 
 

\section*{DATA AVAILABILITY}

The X-ray data underlying this article were retrieved from the public archive of {\it XMM-Newton}. The radiative-transfer models produced with {\sc skirt} will be shared on reasonable request to the corresponding author.


\begin{thebibliography}{99}


\bibitem[Arnaud(1996)]{Arnaud1996} Arnaud, K.~A.\ 1996, Astronomical Data Analysis Software and Systems V, 101, 17

\bibitem[Camps \& Baes(2020)]{Camp2020} Camps, P. \& Baes, M.\ 2020, Astronomy and Computing, 31, 100381

\bibitem[Carciofi et al.(2009)]{Carciofi2009} Carciofi, A.~C., Okazaki, A.~T., Le Bouquin, J.-B., et al.\ 2009, \aap, 504, 3, 915

\bibitem[Chomiuk et al.(2021)]{Chomiuk2021} Chomiuk, L., Metzger, B.~D., \& Shen, K.~J.\ 2021, \araa, 59, 391

\bibitem[Cochetti et al.(2019)]{Cochetti2019} Cochetti, Y.~R., Arcos, C., Kanaan, S., et al.\ 2019, \aap, 621, A123

\bibitem[Coe et al.(2020)]{Coe2020} Coe, M.~J., Kennea, J.~A., Evans, P.~A., et al.\ 2020, \mnras, 497, L50

\bibitem[Diplas \& Savage(1994)]{Diplas1994} Diplas, A. and Savage, B. D.\ \apjs, 93, 211

\bibitem[Eze(2014)]{Eze2014} Eze, R.~N.~C.\ 2014, \mnras, 437, 857

\bibitem[Franchini \& Martin(2021)]{Franchini2021} Franchini, A. \& Martin, R.~G.\ 2021, \apjl, 923, L18

\bibitem[Frank et al.(2002)]{Frank2002} Frank, J., King, A., \& Raine, D.~J.\ 2002, Accretion Power in Astrophysics, by Juhan Frank and Andrew King and Derek Raine, pp. 398. ISBN 0521620538. Cambridge, UK: Cambridge University Press, February 2002., 398

\bibitem[Gabriel et al.(2004)]{Gabriel2004} Gabriel, C., Denby, M., Fyfe, D.~J., et al.\ 2004, Astronomical Data Analysis Software and Systems (ADASS) XIII, 314, 759

\bibitem[Gies et al.(2023)]{Gies2023} Gies, D.~R., Wang, L., \& Klement, R.\ 2023, \apjl, 942, L6

\bibitem[Gies et al.(2007)]{Gies2007} Gies, D.~R., Bagnuolo, W.~G., Baines, E.~K., et al.\ 2007, \apj, 654, 1, 527

\bibitem[Guerrero et al.(2024)]{Guerrero2024} Guerrero, M.~A., Montez, R., Ortiz, R., et al.\ 2024, \aap, 689, A62

\bibitem[Gunderson et al.(2025)]{Gunderson2025} Gunderson, S.~J., Huenemoerder, D.~P., Torrej{\'o}n, J.~M., et al.\ 2025, \apj, 978, 1, 105

\bibitem[Haberl(1995)]{Haberl1995} Haberl, F.\ 1995, \aap, 296, 685. 

\bibitem[Hamaguchi et al.(2016)]{Hamaguchi2016} Hamaguchi, K., Oskinova, L., Russell, C.~M.~P., et al.\ 2016, \apj, 832, 140

\bibitem[Hovis-Afflerbach et al.(2025)]{Hovis-Afflerbach2025}
Hovis-Afflerbach, B., G{\"o}tberg, Y., Schootemeijer, A., et al.\ 2025, \aap, 697, A239

\bibitem[Ishida et al.(2009)]{Ishida2009} Ishida, M., Okada, S., Hayashi, T., et al.\ 2009, \pasj, 61, S77

\bibitem[Kahabka et al.(2006)]{Kahabka2006} Kahabka, P., Haberl, F., Payne, J.~L., et al.\ 2006, \aap, 458, 285

\bibitem[Kennea et al.(2021)]{Kennea2021} Kennea, J.~A., Coe, M.~J., Evans, P.~A., et al.\ 2021, \mnras, 508, 781

\bibitem[Kubo et al.(1998)]{Kubo1998} Kubo, S., Murakami, T., Ishida, M., et al.\ 1998, \pasj, 50, 417 

\bibitem[Langer et al.(2020)]{Langer2020} Langer, N., Baade, D., Bodensteiner, J., et al.\ 2020, \aap, 633, A40

\bibitem[Li et al.(2012)]{Li2012} Li, K.~L., Kong, A.~K.~H., Charles, P.~A., et al.\ 2012, \apj, 761, 99

\bibitem[Lima et al.(2024)]{Lima2024} Lima, I.~J., Luna, G.~J.~M., Mukai, K., et al.\ 2024, \aap, 689, A86

\bibitem[Lodders et al.(2009)]{Lodders2009} Lodders, K., Palme, H., \& Gail, H.-P.\ 2009, Landolt B{\"o}rnstein, 4B, 712

\bibitem[Lucy et al.(2020)]{Lucy2020} Lucy, A.~B., Sokoloski, J.~L., Munari, U., et al.\ 2020, \mnras, 492, 3107

\bibitem[Luna et al.(2013)]{Luna2013} Luna, G.~J.~M., Sokoloski, J.~L., Mukai, K., et al.\ 2013, \aap, 559, A6

\bibitem[MacLeod \& Loeb(2023)]{MacLeod2023} MacLeod, M. \& Loeb, A.\ 2023, Nature Astronomy, 7, 1218

\bibitem[Magdziarz \& Zdziarski(1995)]{Magdziarz1995} Magdziarz, P. \& Zdziarski, A.~A.\ 1995, \mnras, 273, 3, 837 

\bibitem[Martin et al.(2024)]{Martin2024} Martin, R.~G., Lubow, S.~H., Armitage, P.~J., et al.\ 2024, \mnras, 530, 4, 4148 

\bibitem[Martin et al.(2014)]{Martin2014} Martin, R.~G., Nixon, C., Armitage, P.~J., et al.\ 2014, \apjl, 790, L34

\bibitem[Merc(2025)]{Merc2025} Merc, J.\ 2025, Galaxies, 13, 3, 49

\bibitem[Mukai(2017)]{Mukai2017} Mukai, K.\ 2017, \pasp, 129, 062001. doi:10.1088/1538-3873/aa6736

\bibitem[Murakami et al.(1986)]{Murakami1986} Murakami, T., Koyama, K., Inoue, H., et al.\ 1986, \apjl, 310, L31

\bibitem[M\"{u}rset et al.(1997)]{Murset1997} M\"{u}rset, U., Wolff, B., \& Jordan, S.\ 1997, \aap, 319, 201

\bibitem[Naz{\'e}(2025)]{Naze2025} Naz{\'e}, Y.\ 2025, Galaxies, 13, 1, 8

\bibitem[Naz{\'e} et al.(2024)]{Naze2024} Naz{\'e}, Y., Motch, C., Rauw, G., et al.\ 2024, \aap, 688, A181

\bibitem[Naz{\'e} et al.(2022)]{Naze2022} Naz{\'e}, Y., Rauw, G., Smith, M.~A., et al.\ 2022, \mnras, 516, 3366

\bibitem[Naz{\'e} \& Robrade(2023)]{Naze2023} Naz{\'e}, Y. \& Robrade, J.\ 2023, \mnras, 525, 4186

\bibitem[Naz{\'e} \& Motch(2018)]{Naze2018} Naz{\'e}, Y. \& Motch, C.\ 2018, \aap, 619, A148

\bibitem[Okazaki et al.(2002)]{Okazaki2002} Okazaki, A.~T., Bate, M.~R., Ogilvie, G.~I., et al.\ 2002, \mnras, 337, 967

\bibitem[Orio et al.(2007)]{Orio2007} Orio, M., Zezas, A., Munari, U., et al.\ 2007, \apj, 661, 1105

\bibitem[Oskinova et al.(2011)]{Oskinova2011} Oskinova, L.~M., Todt, H., Ignace, R., et al.\ 2011, \mnras, 416, 2, 1456

\bibitem[Patterson \& Raymond(1985)]{Patterson1985} Patterson, J. \& Raymond, J.~C.\ 1985, \apj, 292, 535

\bibitem[Postnov et al.(2017)]{Postnov2017} Postnov, K., Oskinova, L., \& Torrej{\'o}n, J.~M.\ 2017, \mnras, 465, L119

\bibitem[Pringle \& Savonije(1979)]{Pringle1979} Pringle, J.~E. \& Savonije, G.~J.\ 1979, \mnras, 187, 777

\bibitem[Raguzova(2001)]{Raguzova2001} Raguzova, N.~V.\ 2001, \aap, 367, 848

\bibitem[Rast et al.(2025)]{Rast2025} Rast, R.~G., Jones, C.~E., Suffak, M.~W., et al.\ 2025, \mnras, 537, 4, 3575 

\bibitem[Rauw(2024)]{Rauw2024} Rauw, G.\ 2024, \aap, 682, A179

\bibitem[Robinson \& Smith(2000)]{Robinson2000} Robinson, R.~D. \& Smith, M.~A.\ 2000, \apj, 540, 474

\bibitem[Ru{\v{z}}djak et al.(2009)]{Ruz2009} Ru{\v{z}}djak, D., Bo{\v{z}}i{\'c}, H., Harmanec, P., et al.\ 2009, \aap, 506, 1319

\bibitem[Sacchi et al.(2024)]{Sacchi2024} Sacchi, A., Karovska, M., Raymond, J., et al.\ 2024, \apj, 961, 12

\bibitem[Shakura \& Sunyaev(1973)]{Shakura1973} Shakura, N.~I. \& Sunyaev, R.~A.\ 1973, \aap, 24, 337. 

\bibitem[Shao \& Li(2021)]{ShaoLi2021} Shao, Y. \& Li, X.-D.\ 2021, \apj,  920, 2, 81

\bibitem[Smith et al.(2001)]{Smith2001} Smith, R.~K., Brickhouse, N.~S., Liedahl, D.~A., et al.\ 2001, \apjl, 556, 2, L91

\bibitem[Smith et al.(2016)]{Smith2016} Smith, M.~A., Lopes de Oliveira, R., \& Motch, C.\ 2016, Advances in Space Research, 58, 782

\bibitem[Smith \& Robinson(1999)]{SmithRobinson1999} Smith, M.~A. \& Robinson, R.~D.\ 1999, \apj, 517, 866

\bibitem[Sturm et al.(2012)]{Sturm2012} Sturm, R., Haberl, F., Pietsch, W., et al.\ 2012, \aap, 537, A76

\bibitem[Tejeda \& Toal{\'a}(2025)]{TejedaToala2025} Tejeda, E. \& Toal{\'a}, J.~A.\ 2025, \apj, 980, 2, 226

\bibitem[Toal{\'a}(2024)]{Toala2024} Toal{\'a}, J.~A.\ 2024, \mnras, 528, 987

\bibitem[Toal{\'a} et al.(2024)]{Toala2024_TCrB} Toal{\'a}, J.~A., Gonz{\'a}lez-Mart{\'\i}n, O., Sacchi, A., et al.\ 2024, \mnras, 532, 1421

\bibitem[Toal{\'a} et al.(2023)]{Toala2023} Toal{\'a}, J.~A., Gonz{\'a}lez-Mart{\'\i}n, O., Karovska, M., et al.\ 2023a, \mnras, 522, 6102

\bibitem[Toal{\'a} et al.(2022)]{Toala2022} Toal{\'a}, J.~A., Sabin, L., Guerrero, M.~A., et al.\ 2022, \apjl, 927, L20

\bibitem[Tsujimoto et al.(2018)]{Tsujimoto2018} Tsujimoto, M., Morihana, K., Hayashi, T., et al.\ 2018, \pasj, 70, 6, 109

\bibitem[Tycner et al.(2004)]{Tycner2004} Tycner, C., Hajian, A.~R., Armstrong, J.~T., et al.\ 2004, \aj, 127, 2, 1194

\bibitem[Quirrenbach et al.(1997)]{Quirrenbach1997} Quirrenbach, A., Bjorkman, K.~S., Bjorkman, J.~E., et al.\ 1997, \apj, 479, 1, 477

\bibitem[Vander Meulen et al.(2023)]{vanderMeulen2023} Vander Meulen, B., Camps, P., Stalevski, M., et al.\ 2023, \aap, 674, A123

\bibitem[Vasquez-Torres et al.(2024)]{VT2024} Vasquez-Torres, D.~A., Toal{\'a}, J.~A., Sacchi, A., et al.\ 2024, \mnras, 535, 3, 2724

\bibitem[Wang et al.(2023)]{Wang2023} Wang, L., Gies, D. R., Peters, G. J., Han, Z.\ 2023, \aj, 165, 203

\bibitem[Wheatley \& Kallman(2006)]{Wheatley2006} Wheatley, P.~J. \& Kallman, T.~R.\ 2006, \mnras, 372, 1602

\bibitem[White et al.(1982)]{White1982} White, N.~E., Swank, J.~H., Holt, S.~S., et al.\ 1982, \apj, 263, 277

\bibitem[Wilms et al.(2000)]{Wilms2000} Wilms, J., Allen, A., \& McCray, R.\ 2000, \apj, 542, 914

\bibitem[Yungelson et al.(2024)]{Yungelson2024} Yungelson, L., Kuranov, A., Postnov, K., et al.\ 2024, \aap, 2024, 683, A37 

\bibitem[Zhekov \& Tomov(2019)]{Zhekov2019} Zhekov, S.~A. \& Tomov, T.~V.\ 2019, \mnras, 489, 2930

\bibitem[Zhu et al.(2023)]{Zhu2023} Zhu, C.-H., L{\"u}, G.-L., Lu, X.-Z., et al.\ 2023, Research in Astronomy and Astrophysics, 23, 2, 025021

\end{thebibliography}
\end{document}